\documentclass[prl,nofootinbib,twocolumn,showpacs,amssymb]{revtex4}

\usepackage{amsmath}
\usepackage{amsfonts}
\usepackage{amssymb}
\usepackage{color,graphicx,shortvrb,epsfig}
\usepackage{latexsym}
\usepackage[a4paper]{geometry}
\usepackage{hyperref}
\usepackage[in]{fullpage}

\begin{document}

\title{Flavoring Astrophysical Neutrinos: Flavor Ratios Depend on Energy}

\author{Tamar Kashti \& Eli Waxman}
\affiliation{Physics Faculty, Weizmann Institute of Science, Rehovot 76100, Israel}

\begin{abstract}
Electromagnetic (and adiabatic) energy losses of $\pi$'s and
$\mu$'s modify the flavor ratio (measured at Earth) of neutrinos
produced by $\pi$ decay in astrophysical sources,
$\Phi_{\nu_e}:\Phi_{\nu_\mu}:\Phi_{\nu_\tau}$, from $1:1:1$ at low
energy to $1:1.8:1.8$ at high energy. The transition occurs over
1--2 decades of $\nu$ energy, and is correlated with a
modification of the neutrino spectrum. For $\gamma$-ray bursts,
e.g., the transition is expected at $\sim100$~TeV, and may be
detected by km-scale $\nu$ telescopes. Measurements of the
transition energy and energy-width will provide unique probes of
the physics of the sources. $\pi$ and $\mu$ energy losses also
affect the ratio of $\bar\nu_e$ flux to total $\nu$ flux, which
may be measured at the $W$-resonance (6.3~PeV): It is modified
from $1/6$ ($1/15$) at low energy to $1/9$ (practically 0) at high
energy for neutrinos produced in $pp$ ($p\gamma$) interactions.
\end{abstract}

\maketitle

The existence of extra Galactic high energy neutrino
sources (for reviews see \cite{Gaisser:2003sc}) is implied by
observations of ultra-high energy cosmic rays with energies of
$\epsilon>10^{19}$~eV. The acceleration of particles in extra-Galactic
sources is expected to produce a cosmic-ray spectrum extending over many decades
of energy (although the observed spectrums is likely dominated by
Galactic sources below $\sim10^{19}$~eV), leading to the production of
a wide energy spectrum of extra-Galactic high energy neutrinos.
Possible sources of high energy neutrinos
include $\gamma$-ray bursts (GRBs) and active galactic nuclei
(AGNs). High energy ($>0.1$~TeV) neutrino telescopes are being
constructed in order to detect extra Galactic neutrinos (for
review see \cite{Halzen:2005wd}). Their detection will allow one
to identify the high energy cosmic ray sources, and to probe their
physics. It may also provide information on fundamental neutrino
properties.

High energy neutrinos are believed to be produced in astrophysical
sources mainly through the decay of charged pions, $\pi^+\to\mu^+
+\nu_\mu\to e^+ + \nu_\mu + \nu_e + \bar \nu_\mu$ or $\pi^-\to\mu^-
+\bar\nu_\mu\to e^- + \nu_\mu + \bar\nu_e + \bar \nu_\mu$, produced
in interactions of high energy protons with "target" photons
($p\gamma$) or nucleons ($pp$, $pn$)\footnote{$\nu_\tau$'s may also be
produced, by production of charmed mesons, but the higher energy
threshold and lower cross-section for charmed meson production typically
imply a negligible $\nu_\tau$ fraction, e.g. \cite{Waxman:1997ti}}.
The ratio of the fluxes of
neutrinos of different flavors is therefore expected to be, at the
source, $\Phi_{\nu_e}^0:\Phi_{\nu_\mu}^0:\Phi_{\nu_\tau}^0=1:2:0$
($\Phi_{\nu_l}$ stands for the combined flux of $\nu_l$ and
$\bar\nu_l$). Neutrino oscillations then lead to an observed
flux ratio on Earth of
$\Phi_{\nu_e}:\Phi_{\nu_\mu}:\Phi_{\nu_\tau}=1:1:1$
\cite{Learned:1994wg}. For neutrinos produced in $pp$ ($pn$)
collisions, where both $\pi^+$'s and $\pi^-$'s are produced, the
ratio of $\bar\nu_e$ flux to total $\nu$ flux is $1/6$, while for
neutrinos produced in $p\gamma$ collisions, where only $\pi^+$'s are
produced, the ratio is $1/15$
\cite{Ackermann:2004pj,Anchordoqui:2004eb}.

Km-scale optical Cerenkov neutrino telescopes, such as IceCube, are
expected to be capable of discriminating between neutrino flavors
\cite{Beacom:2003nh}. Moreover, although these detectors can not distinguish
between neutrinos and anti-neutrinos at most energies, they may
allow one to determine the ratio of $\nu_e$ and $\bar\nu_e$ fluxes
by detecting $\bar\nu_e$'s at the $W$-resonance ($\bar\nu_e e^-\to
W^-\to$ anything) around $6.3$~PeV (see, e.g., \cite{Anchordoqui:2004eb}
for discussion of resonant $\bar\nu_e$ detection in IceCube). It has been pointed out
\cite{Learned:1994wg,Athar:2000yw} that measurements of the flavor
composition of astrophysical high energy neutrinos may enable one to
probe new physics, by searching for deviations from the standard
flavor ratio $1:1:1$. Measurements of the $\nu_e$ to $\bar\nu_e$
flux ratio may allow one to probe the physics of the sources
\cite{Anchordoqui:2004eb}, by discriminating between the two primary
modes of pion production, $p\gamma$ and $pp$ collisions.

Although a $1:1:1$ flavor ratio appears to be a robust prediction of
models where neutrinos are produced by pion decay, we point out in
this letter that energy dependence of the flavor ratio is a generic
feature of models of high energy astrophysical neutrino sources.
Pions are typically produced in environments where they may suffer
significant energy losses prior to decay, due to interaction with
radiation and magnetic fields \cite{Waxman:1997ti,Rachen:1998fd}.
Since the pion life time is shorter than the muon's, at sufficiently
high energy the probability for pion decay prior to significant
energy loss is higher than the corresponding probability for muon
decay. This leads to suppression at high energy of the relative
contribution of muon decay to the neutrino flux. The flavor ratio is
modified to $0:1:0$ at the source, similar to that of atmospheric
neutrinos at high energies where the muons do not decay (see
e.g.~\cite{Gonzalez-Garcia:2002dz}), implying $1:1.8:1.8$ ratio on
Earth \cite{Beacom:2003nh}.

The energy dependence of the neutrino flavor content provides a
unique probe of the sources. On the other hand, it complicates
attempts to study new physics based on measuring deviations from
$\Phi_{\nu_e}:\Phi_{\nu_\mu}=1:1$. Furthermore, if muon energy
losses become important at $\le1$~PeV, it would affect the flux of
$\bar\nu_e$ near the W resonance, rendering the suggestion to probe
neutrino mixing angles with neutrinos around the W resonance
\cite{Bhattacharjee:2005nh} impractical, and making the
discrimination between $p\gamma$ and $pp$ collisions more difficult
(due to reduction of $\Phi_{\bar\nu_e}:\Phi_\nu^{\rm total}$).

We first derive below approximate analytic expressions describing
the energy dependence of the neutrino flux, flavor ratio and
anti-particle content, for sources that produce pions with a
power-law energy distribution [a differential number flux
$\Phi_\pi(\epsilon_\pi)\propto\epsilon_\pi^{-k}]$, and assuming
the charged particle energy loss rate to be proportional to a
power of the particle energy, $\dot{\epsilon}\propto-\epsilon^n$.
Such energy dependence is expected for synchrotron and
inverse-Compton emission (below the Klein-Nishina regime), in
which case $\dot{\epsilon}\propto-\epsilon^2$, and for "adiabatic"
energy loss (energy loss due to expansion of the plasma in which
the particles are confined), in which case
$\dot{\epsilon}\propto-\epsilon^1$. We then discuss several
specific models of high-energy neutrino sources, and the general
implications of our results.

\noindent{\it The energy dependence of the flavor ratio.}
\label{sec:energy} We consider neutrinos from astrophysical
sources produced by the decay of charged pions. The decay of the
pion and the subsequent decay of the muon lead to a flavor ratio
of $\Phi_{\nu_e}^0:\Phi_{\nu_\mu}^0:\Phi_{\nu_\tau}^0=1:2:0$. We
note that in the production of a charged pion, the high energy
proton may be converted to a high energy neutron (e.g. $p+\gamma\to
n+\pi^+$). The neutron may escape the source and decay, producing
an additional $\bar\nu_e$. However, in this decay the neutrino
carries only a small fraction, $\sim10^{-3}$, of the original proton
energy, comparable to $(m_n-m_p)/m_n$, much lower than the energy of
neutrinos produced by the pion decay. Since in most astrophysical
sources the number of emitted neutrinos drops rapidly with energy,
the contribution of neutron decay to the neutrino flux would
generally be small and is therefore neglected here.

The flavor ratio
$\Phi_{\nu_e}^0:\Phi_{\nu_\mu}^0:\Phi_{\nu_\tau}^0=1:2:0$  is
modified at high energy, where the life time of the muons becomes
comparable to, or smaller than, the time for significant
electromagnetic (or adiabatic) energy loss. The particle life
time, $\tau_{x,\rm decay}$ were $x$ stands for $\pi^\pm$ or
$\mu^\pm$, is proportional to $\epsilon_x$. Thus, the ratio of
cooling time, $\tau_{x,\rm
cool}\equiv\epsilon_x/|\dot{\epsilon}_x|$, to life time is rapidly
decreasing with energy, $\tau_{x,\rm cool}/\tau_{x,\rm
decay}\propto\epsilon_x^{-n}$. We denote the energy at which
$\tau_{x,\rm cool}=\tau_{x,\rm decay}$ by $\epsilon_{0,x}$. The
$1/\epsilon_x^n$ dependence of the cooling and decay time ratio
implies that $\epsilon_{0,\pi}/\epsilon_{0,\mu}$ is approximately
given by $(\tau_{0,\pi}/\tau_{0,\mu})^{-1/n}\sim 10^{2/n}$, where
$\tau_{0,\pi}=2.6\times 10^{-8}$~s and $\tau_{0,\mu}=2.2\times
10^{-6}$~s are the pion and muon rest-frame life times.

We consider below neutrinos produced by the decay of $\pi^+$'s. The
results for neutrinos from $\pi^-$ decay are simply given by
replacing each particle with its anti-particle. For pion decay, each
of the four final leptons carry approximately $1/4$ of the pion
energy. The (differential number) flux of $\nu_\mu$'s of energy
$\epsilon_\nu$ produced by $\pi^+$ decay is therefore approximately
given by
\begin{align}\label{n_nu}
    \Phi^0_{\nu_\mu}(\epsilon_\nu)
    =-\partial_{\epsilon_\nu}\int_{4\epsilon_\nu}^\infty d\epsilon_i
    \Phi_\pi(\epsilon_i)P_\pi(\epsilon_{i},4\epsilon_\nu).
\end{align}
Here, $\Phi_\pi(\epsilon_i)$ is number flux (per unit energy) of
pions produced by the source with energy $\epsilon_i$, and
$P_\pi(\epsilon_{i},4\epsilon_\nu)$ is the probability that a pion
produced with energy $\epsilon_i$ would decay before its energy
drops below $4\epsilon_\nu$.

For $\dot{\epsilon}\propto\epsilon^n$ with $n>0$,
$P(\epsilon_i,\epsilon_f)=1-\exp[-\epsilon_0^n(\epsilon_f^{-n}-\epsilon_i^{-n})/n]$.
Assuming $\Phi_\pi(\epsilon_{i})\propto\epsilon_{i}^{-k}$ with
$k>1$, we find that the neutrino flux is suppressed due to energy
loss by a factor
\begin{align}\label{flux from pi}
    \frac{\Phi^0_{\nu_\mu}(\epsilon_\nu)}{\Phi_0^{(\rm no\mbox{ }loss)}}
    &= s(-s)^{\frac{1-k}{n}} e^{-s}
    \gamma\left(\frac{k-1}{n},-s\right).
\end{align}
Here $s\equiv(\epsilon_{0,\pi}/4\epsilon_\nu)^n/n$, $\gamma(a,z)$
is the lower incomplete gamma function, and $\Phi_0^{(\rm no\mbox{
}loss)}$ is the flux that would have been obtained had energy
losses been negligible.

\begin{figure*}[t]
   \centerline{\includegraphics[width=0.95\textwidth]{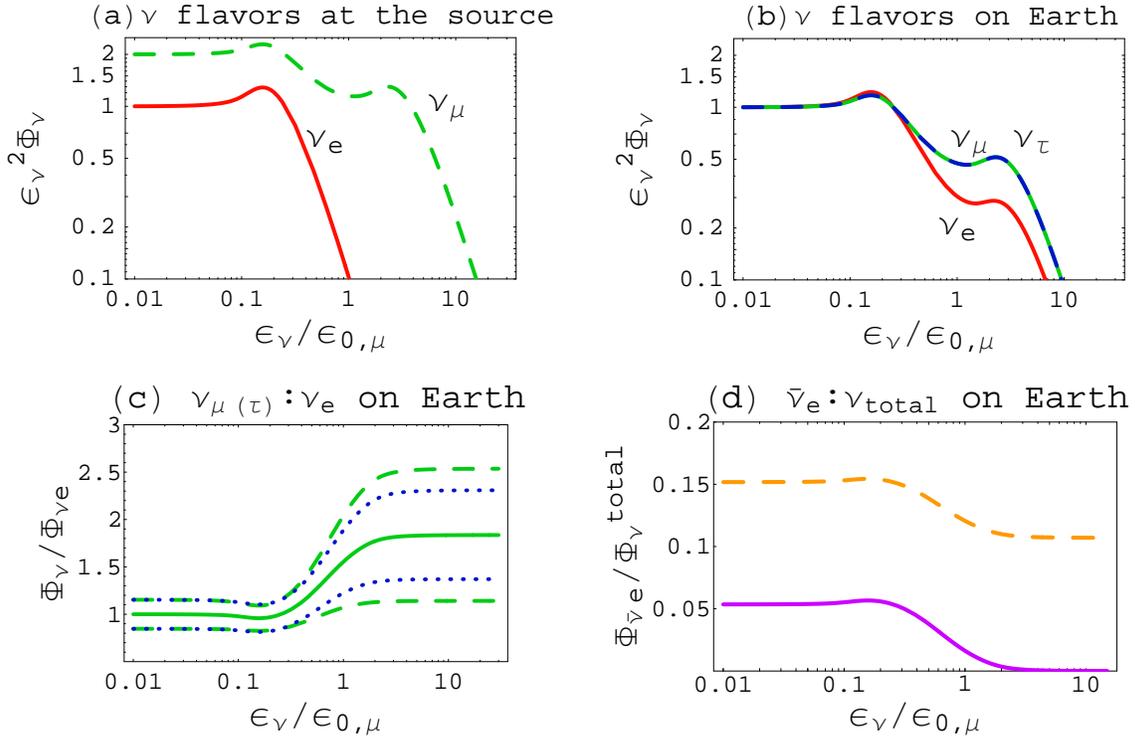}}
    \caption{Flavor and anti-particle content of the flux of
    astrophysical neutrinos produced by pion decay, for $\Phi_\pi\propto1/\epsilon_\pi^{2}$
    and $\dot{\epsilon}_x\propto\epsilon_x^2$. Figures (a-b) present the energy fluxes in different flavors,
    $\epsilon_{\nu_l}^2\Phi_{\nu_l}$ (normalized to $\epsilon_{\nu_e}^2\Phi_{\nu_e}$ at low energy).
    $\Phi_{\nu_l}$ stands for the combined flux of $\nu_l$ and $\bar\nu_l$,
    and these plots are therefore valid for neutrinos produced by any combination of $\pi^+$ and $\pi^-$ decay.
    Figure (c) presents the ratio between $\Phi_{\nu_{\mu}(\nu_\tau)}$
    and $\Phi_{\nu_e}$ (solid line), with 90\% CL lines of $\nu_\mu$ (dashed) and $\nu_\tau$ (dotted) fluxes.
    Figure (d) presents the ratio of $\bar\nu_e$ to
     total $\nu$ flux on Earth, solid (dashed) line for neutrinos produced by  $p\gamma$ ($pp$) interactions.}
    \label{fig:Nnu}
\end{figure*}

Similarly, for $\nu$'s produced by the decay of $\mu^+$ we have
\begin{align}
\nonumber
  \Phi^0_{\bar\nu_\mu(\nu_e)}(\epsilon_\nu)=&\partial_{\epsilon_\nu}\int_{3\epsilon_\nu}^\infty
    d\epsilon_\mu \int_{\frac43\epsilon_\mu}^\infty d\epsilon_i\Phi_\pi
\\ &\times P_\mu(\!\epsilon_\mu,3\epsilon_\nu)
   \partial_{\epsilon_\mu}P_\pi(\epsilon_{i},4\epsilon_\mu/3),
\end{align}
and
\begin{align}\label{flux from mu}
&\frac{\Phi^0_{\bar\nu_\mu(\nu_e)}(\epsilon_\nu)}{\Phi_0^{(\rm no\mbox{ }loss)}}
    =\frac{q^n}{1-q^n}(-s)^{\frac{1-k+n}{n}}\\
    & \!\times\! \left[e^{-s}\gamma\!\left(\frac{k-1}{n},-s\right)\!\!
     -e^{-q^n \!s}q^{1-k} \gamma\!\left(\frac{k-1}{n},\!-q^n
     s\right)\right], \nonumber
\end{align}
where $q\equiv4\epsilon_{0,\mu}/3\epsilon_{0,\pi}$.

The flavor content given in eqs.~(\ref{flux from pi},\ref{flux
from mu}) is modified as the $\nu$'s propagate from the source to
Earth. For propagation over cosmological distances $d$,
$d\gg \hbar c\epsilon/\Delta m^2c^4$ and the observed fluxes
$\Phi_{\nu_\alpha}$ ($\alpha={e,\mu,\tau}$) are related to the
production fluxes, $\Phi^0_{\nu_\alpha}$, by (see e.g.
\cite{Gonzalez-Garcia:2002dz})
\begin{align}
\label{eq:mixing}
    \Phi_{\nu_\alpha}\!=\!\sum_\beta P_{\alpha\beta}\Phi^0_{\nu_\beta}
    \!=\!\sum_\beta \sum_i \left|U_{\alpha i}\right|^2 \left|U_{\beta i}\right|^2\!\Phi^0_{\nu_\beta}.
\end{align}
Here, $U_{\alpha i}$ is the neutrino mixing matrix, $
|\nu_\alpha\rangle =\sum_{i=1}^3 U_{\alpha i}^*|\nu_i\rangle$
where $\nu_i$ ($i={1,2,3}$) are the mass eigen-states.

$U_{\alpha i}$ can be written as a function of three mixing angles
$\theta_{12}, \theta_{23}, \theta_{13}$ and a Dirac phase
$\delta$. The best fit for $\delta=0$ \cite{Strumia:2005tc} is:
$\theta_{12}=34^o\pm2.5^o$, $\theta_{23}=45^o\pm6^o$,
$\theta_{13}=0^o\pm8^o$ (90\% confidence level). The observed
neutrino flux ratio is
$\Phi_{\nu_e}:\Phi_{\nu_\mu}:\Phi_{\nu_\tau}=1:1.00\pm0.15:1.00\pm0.15$
for initial flux ratio
$\Phi_{\nu_e}^0:\Phi_{\nu_\mu}^0:\Phi_{\nu_\tau}^0=1:2:0$, and
$\Phi_{\nu_e}:\Phi_{\nu_\mu}:\Phi_{\nu_\tau}=1:1.8^\pm0.70:1.8\pm0.47$
for $\Phi_{\nu_e}^0:\Phi_{\nu_\mu}^0:\Phi_{\nu_\tau}^0=0:1:0$
(90\% confidence level). Here, we have obtained an approximate
estimate of the error bars by assuming the errors in different
mixing angles are not correlated, i.e.
$\Delta(\Phi_{\nu_\alpha}/\Phi_{\nu_e})=\{\sum_{ij}\Delta
\theta_{ij}^2[\partial_{\theta_{ij}}(\Phi_{\nu_\alpha}/\Phi_{\nu_e})]^2\}^{1/2}$.

The results of applying eq.~(\ref{eq:mixing}) to the initial flux
ratios given  by eqs.~(\ref{flux from pi}, \ref{flux from mu}) are
shown in fig.~\ref{fig:Nnu}(a-c), for the case of
$\Phi_\pi\propto1/\epsilon_\pi^{2}$ and
$\dot{\epsilon}_x\propto\epsilon_x^2$. As expected, the muon
(pion) decay neutrino flux is suppressed above
$\epsilon_\nu\sim\epsilon_{0,\mu}/3$
($\epsilon_\nu\sim\epsilon_{0,\pi}/4$) by a factor
$\propto\epsilon_\nu^{-n}=\epsilon_\nu^{-2}$, reflecting the fact
that at high energy the probability for decay prior to significant
energy loss is $\propto\epsilon^{-n}=\epsilon_\nu^{-2}$, and the
flavor ratio transition takes place over an energy range of
$\sim\epsilon_{0,\pi}/\epsilon_{0,\mu}\sim(\tau_{0,\pi}/\tau_{0,\mu})^{-1/n}
\sim 10^{2/n}\sim10$.

The energy at which a flavor ratio transition takes place (from
$1:1:1$ to $1:1.8:1.8$),  and the width of the transition, provide
unique handles on the properties of the source. $\epsilon_{0,\mu}$
may constrain, e.g., the (radiation and magnetic field) energy
density at the source, and the width of the transition may
discriminate between adiabatic and electromagnetic energy losses.
It should be kept in mind, however, that while the flavor ratio
transition is expected to be a generic feature of high energy
neutrinos produced in astrophysical sources, the behavior in the
transition region may be more complicated that described by
eqs.~(\ref{flux from pi}, \ref{flux from mu}), which were obtained
under idealized assumptions. For example, neutrinos may be emitted
from different regions (in a single source) with different values
of $\epsilon_{0,\mu}$ (thus "smearing" the transition), and
suppression of the inverse-Compton scattering cross section in the
Klein-Nishina regime may lead to deviations from a simple
power-law energy dependence of the energy loss rate.

\noindent{\it Neutrinos from specific sources.} \label{sec:GRB}
GRBs are possible sources of high energy neutrinos. The
$\gamma$-rays are believed to be produced by the dissipation of
the kinetic energy of a highly-relativistic wind. Neutrinos that
are expected to be produced in the same region where the
$\gamma$-rays are produced have characteristic energies
$\ge100$~TeV \cite{Waxman:1997ti}. The pions and muons cool by
synchrotron radiation, $\dot{\epsilon}_x\propto\epsilon_x^2$, and
the muon cooling energy in the $\gamma$-ray %\nopagebreak[3]
production region is \cite{Rachen:1998fd,Waxman:1998yy}
\begin{align}\label{cooling energy}
    &\epsilon_{0,\mu}\approx 10^3\frac{\Gamma_{2.5}^4\Delta t_{-3}}{L_{53}^{1/2}}\mbox{TeV}.
\end{align}
Here $\Gamma=10^{2.5}\Gamma_{2.5}$ is the wind Lorentz factor,
$L=10^{53}L_{53}$~erg/s is the kinetic energy luminosity of the
wind (assuming spherically symmetric wind), and $\Delta
t=10^{-3}\Delta t_{-3}$~s is the observed variability time scale
of the $\gamma$-ray signal.

If GRBs are associated, as commonly believed, with collapses of
massive stars, neutrinos of lower energy, $\epsilon_\nu\ge1$~TeV,
may be emitted in the early stage of GRB evolution, when the
relativistic wind penetrates the progenitor star
\cite{Meszaros:2001ms}. The high energy density of radiation
implies, in this scenario, $\epsilon_{0,\mu}<1$~TeV due to
inverse-Compton losses, and the Klein-Nishina suppression of
inverse-Compton losses at high energy implies
$\epsilon_{0,\pi}\gg1$~TeV. In this case, an observed flavor ratio
of $1:1.8:1.8$ would therefore be expected at all energies
($>1$~TeV).

Measurements of the energy dependence of the neutrino flavor ratio
would therefore provide constraints on the physical parameters of
the source, and may allow one to discriminate between different
scenarios for neutrino production in GRBs.

Eq.~(\ref{cooling energy}) holds also for neutrinos produced in
blazar AGN jets \cite{Rachen:1998fd}. For the parameters
characterizing these objects, $\Gamma\sim10$, $L\sim10^{47}$~erg/s
and $\Delta t\sim10^{4}$~s, we have
$\epsilon_{0,\mu}\sim4\times10^{6}$~TeV, implying a flavor transition
at $\sim 10^6$~TeV. When the dominant cooling process is
adiabatic cooling, a similar cooling energy is obtained. However, at
these energies, the number of neutrinos detected from AGNs may be
too small to allow detection of the transition.

\noindent{\it Implications for the flux of $\bar\nu_e$.} The
fraction of $\bar\nu_e$ flux out of the total flux decreases at
high energy due to muon cooling. For neutrinos produced in $pp$
interactions, where $\pi^+$'s  and $\pi^-$'s are produced at
roughly equal numbers, a flavor ratio of $1:1:1$ is obtained at
low energy for both $\nu_\alpha$ and $\bar\nu_\alpha$, implying
$\Phi_{\bar\nu_e}:\Phi_\nu^{\rm total}=1:6$. At high energy, the
flavor ratio changes to $1:1.8:1.8$, implying
$\Phi_{\bar\nu_e}:\Phi_\nu^{\rm total}=1:9$. For neutrinos
produced in $p\gamma$ interactions, where only $\pi^+$'s are
created, $\Phi_{\bar\nu_e}:\Phi_\nu^{\rm total}=1:15$ at low
energy. At high energy, the muon energy losses suppress the
production of anti-neutrinos, resulting in a strong suppression of
$\Phi_{\bar\nu_e}/\Phi_\nu^{\rm total}$. This is illustrated in
fig.~\ref{fig:Nnu}(d), where the energy dependence of the
$\bar\nu_e$ to $\nu_e$ flux ratio is plotted for
$\Phi_\pi\propto1/\epsilon_\pi^{2}$ and
$\dot{\epsilon}_x\propto\epsilon_x^2$.

If the transition is below 1~PeV, as expected for GRBs,  it will
reduce the flux of $\bar\nu_e$ near the W resonance, making the
detection of $\bar\nu_e$, and hence the discrimination between
$p\gamma$ and $pp$ collisions, more difficult. It should be
pointed out in this context, that significant production of
$\pi^-$'s may occur in sources where pions are produced only
through interactions of nucleons with photons, if the
photo-production optical depth is large enough, so that the
neutron produced in a $p\gamma\to n+\pi^+$ interaction is likely
to interact with a photon before escaping the source ($n\gamma\to
p+\pi^-$). Thus, a discrimination between photo-production and
inelastic nuclear collision sources based on the observed
$\Phi_{\bar\nu_e}/\Phi_\nu^{\rm total}$ ratio may not be straight
forward.

\noindent{\it Conclusions.} We have shown that a flavor ratio
transition, from
$\Phi_{\nu_e}:\Phi_{\nu_\mu}:\Phi_{\nu_\tau}=1:1:1$ at low energy
to $1:1.8:1.8$ at high energy, is expected to be a generic feature
of high energy neutrinos produced in astrophysical sources. The
location and energy width of the transition provide unique handles
on the properties of the source, and may allow one to discriminate
between different scenarios for neutrino production. The modified
flavor ratio affects the experimental upper limits on the total
neutrino flux, which are commonly obtained assuming a $1:1:1$
ratio (e.g. \cite{Ackermann:2004pj}). It also changes the ratio of
$\bar\nu_e$ to total $\nu$ flux from $1/6$ ($1/15$) at low energy
to $1/9$ (practically 0) at high energy for neutrinos produced in
$pp$ ($p\gamma$) interactions.

We have neglected in the current analysis the possibility of
matter oscillations, which are typically unimportant for the high
energy sources under consideration (e.g.~\cite{Waxman:1997ti}).
However, it should be kept in mind that matter oscillations may,
in general, play a role in modifying the predicted flavor ratio.
For example, the flavor ratio of neutrinos produced by a GRB jet
buried in a blue super giant would be modified by
Mikheyev-Smirnov-Wolfenstein resonance to $1:1.5:1.5$ outside the
source and $1.2:1.4:1.4$ on Earth.

\noindent{\it Acknowledgments.} We would like to thank
C.~Pe\~{n}a-Garay, Y.~Nir \& B.~Katz for helpful discussions. This
research was supported in part by ISF and Minerva grants.
%\vspace{-0.5cm}

\end{document}